\documentclass{article}

\usepackage[preprint]{neurips_2022}
\usepackage{graphicx}
\usepackage[utf8]{inputenc} 
\usepackage[T1]{fontenc}    
\usepackage{hyperref}       
\usepackage{url}            
\usepackage{booktabs}       
\usepackage{amsfonts}       
\usepackage{nicefrac}       
\usepackage{microtype}      
\usepackage{xcolor}         
\usepackage{amsmath}
\usepackage{amssymb}

\title{Emergent Resource Exchange and Tolerated Theft Behavior using Multi-Agent Reinforcement Learning}

\author{
    Jack Garbus \affil{1}, 
    Jordan Pollack \affil{1},
}

\author{%
  \vspace{2.5mm}
  Jack Garbus \hspace{6mm} Jordan Pollack \hspace{6mm} \\
  Department of Computer Science, Brandeis University\\Waltham, Massachussetts \\
  \vspace{3mm}
  \texttt{garbus@brandeis.edu} \\
}

\begin{document}
\maketitle
\begin{abstract}
For decades, the evolution of cooperation has piqued the interest of numerous academic disciplines such as game theory, economics, biology, and computer science. In this work, we demonstrate the emergence of a novel and effective resource exchange protocol formed by dropping and picking up resources in a foraging environment. This form of cooperation is made possible by the introduction of a campfire, which adds an extended period of congregation and downtime for agents to explore otherwise unlikely interactions. We find that the agents learn to avoid getting cheated by their exchange partners, but not always from a third party. We also observe the emergence of behavior analogous to tolerated theft, despite the lack of any punishment, combat, or larceny mechanism in the environment. 
\end{abstract}

\hypertarget{introduction}{%
\section{Introduction}\label{introduction}}

While many consider human intelligence a core factor in the success of our
species, perhaps an even more fundamental component is our ability to
cooperate with one another \citep{henrich_secret_2016}. It is no
surprise then that the emergence of cooperative behavior has long been
an area of interdisciplinary study as researchers seek to model societal
dynamics \citep{epstein_growing_1996,gostoli_self-isolation_2023} and design artificial intelligence
to work with humans \citep{cultural_general_intelligence_team_learning_2022,gauthier_paradigm_2016} .

\citet{axelrod_evolution_1984} demonstrated the power of cooperation through ecological
simulations where a diverse population of strategies faced each other in
the Iterated Prisoner's Dilemma (IPD) game. Cooperative strategies that
punished instances of defection eventually dominated the population,
while defecting strategies proved to be unstable in the long run. Other
experiments on the noisy version of IPD performed mutation
and selection on the strategy population, providing glimpses of
open-ended evolution of cooperation \citep{lindren_evolutionary_1992}.
The resulting theory from these experiments has been broadly applied to
understand the emergence of cooperation between organisms
such as bacteria as well as enemy troops during wartime
\citep{axelrod_evolution_1984}.

Multi-Agent Reinforcement Learning (MARL) has been widely applied as a
tool for studying cooperative behavior between agents which optimize
their behavior to maximize a reward provided by an environment
\citep{hughes_inequity_2018,mckee_deep_2021,agapiou_melting_2023,leibo_malthusian_2019,}. In this paradigm, multiple
agents are placed in an environment designed such that cooperation
between individuals is of greater benefit than competition. Many environments have been developed in pursuit of a variety of emergent
social behaviors such as turn-taking, teaching, resource
sharing, reciprocity and language.
\citep{agapiou_melting_2023,lazaridou_multi-agent_2017,gupta_dynamic_2021}.
The success of multi-agent reinforcement learning has led some
researchers to outline a path towards artificial general intelligence
heavily oriented around reproducing human social intelligence \emph{in
silico} \citep{leibo_autocurricula_2019}. If we view the success of
humanity as a story of cooperation rather than isolated
intelligence \citep{henrich_secret_2016}, then this research path is
quite promising.

Despite the potential of social intelligence,
research on artificial societies has remained limited. While agent-based
models allow researchers to study changes to social behavior, they often employ fixed, human-designed approximations of real-world
dynamics and behaviors
\citep{gostoli_self-isolation_2023,hinsch_effects_2023,epstein_growing_1996}. 
Large language models have recently enabled alternative, flexible forms of social modeling such as \emph{social simulacra}, which simulate community interactions between different personas provided by prompts \citep{park_social_2022,park_generative_2023}. These large models are trained on vast amounts of human-generated data, which plays a large role in the behaviors of the social model, thus limiting the study of how intelligent or optimal behaviors may first arise. 
Systems that optimize agent behavior using reinforcement learning are typically configured to study a specific set of emergent social behaviors and often utilize additional modifications to
the algorithm or environment's mechanics. Some additions include the
training of additional classifiers alongside each policy
\citep{vinitsky_learning_2022}, auxiliary losses
\citep{cultural_general_intelligence_team_learning_2022}, or
behavior-specific mechanisms to enable desired behavior such as trading \citep{johanson_emergent_2022,suarez_specialization_2022}. If we are to realize the vision of emergent artificial societies, we would like to discover simple, general-purpose environmental mechanisms that can induce different social behaviors instead of adding an additional layer of complexity per behavior.

To this end, we demonstrate the emergence of resource exchange without
programming any form of exchange protocol into the algorithm or
environment. While agents have successfully leveraged human-designed exchange systems 
and discovered how to barter, trade stocks, and devise tax policies (learning to game the subsequent tax system as well) \citep{johanson_emergent_2022,pricope_deep_2021,zheng_ai_2020}, 
no prior work to our knowledge has demonstrated emergent resource exchange by picking up and placing
resources in an embodied setting, despite the apparent simplicity of the behavior.
We call our environment The Trading Game, as agents
discover the ability to trade resources by picking up and placing down
foraged resources. 

Food sharing is a prevalent phenomenon among various species, and its
evolution has been a topic of interest for researchers in evolutionary
biology and anthropology \citep{kaplan_food_1985}.  \citet{kaplan_food_1985} reviews several hypotheses that have been proposed to explain
resource-sharing from an evolutionary perspective in non-human species,
including kin selection, tolerated theft, reciprocal altruism, and
cooperative acquisition. The kin selection
hypothesis suggests that sharing resources with close genetic relatives
enhances the fitness of shared genes. The tolerated theft hypothesis
proposes that individuals with ample resources allow those without to
steal from them, as defending the resource is more costly than the
resource's value. The reciprocal altruism hypothesis, previously applied
to the iterated prisoner's dilemma, suggests that reciprocating
cooperative behavior can emerge and be an evolutionarily stable strategy, and the
cooperative acquisition hypothesis proposes that social carnivores hunt
together to increase their chances of catching prey. In the Trading Game
described below, where agents cannot fight, hunt, or reproduce, we see
that reciprocal altruism is the primary driver stabilizing the emergence
of exchange. In addition, we observe the emergence of tolerated theft,
despite the lack of any method for stealing resources or
engaging in combat.

Our contributions are as follows:

\begin{itemize}
    \item We introduce the Trading Game, a simple foraging environment which applies pressure for agents to congregate around a campfire at night.
    \item We demonstrate that agents in our environment can learn to exchange resources using drop/pickup actions during a period of extended congregation, whereas agents from prior research environments could not. 
    \item We demonstrate the emergence of a behavior akin to tolerated theft between agents in our environment, despite the lack of any combat, punishment, or larceny system.
    \item Through an ablation study, we demonstrate how reciprocated resource exchange fails to emerge without sufficient pressure to congregate for extended periods of time.
\end{itemize}

\section{Background}

In this section, we briefly review a few of the most relevant environments used to study emergent cooperation and embodied exchange.

\subsection{Cleanup}

The Cleanup
environment poses a complex social dilemma for multi-agent research, and has
been used to study how systems for reputation, social influence, inequity aversion, and public sanctioning shape emergent cooperation  \citep{mckee_deep_2021,jaques_social_2019,hughes_inequity_2018, vinitsky_learning_2022}.
In Cleanup, agents must simultaneously clear pollution from a river and
collect apples which spawn proportionally to the amount of pollution
cleared. In order to prevent free-riders from collecting apples without
clearing pollution, agents are equipped with a ``punish beam'', which
they can fire at other agents to fine them with a certain amount of negative reward.
This punishment mechanism allows agents to punish free-riders who do not contribute to the cleaning effort. When the punishment beam is combined with one of the additional systems mentioned above, agents learn to punish free-riders to achieve socially beneficial outcomes and overcome the social dilemma. 

\subsection{Fruit Market}
In the Fruit Market environment described in \citet{johanson_emergent_2022}, agents can move around, produce and consume apples or
bananas, and broadcast one of nineteen human-designed trade offers to nearby
agents that are automatically executed by the environment once an agent
accepts. Agents are designated as Apple Farmers (producing more apples
at a time) or Banana Farmers (producing more bananas at a time). When
Apple Farmers receive a larger reward for consuming bananas than apples,
they are incentivized to produce apples and trade them for bananas (and
vice versa for Banana Farmers). Agents eventually converge on trading as
the optimal strategy, and a behavior akin to bartering soon emerges,
where agents broadcast the offer that most benefits them,
lowering their prices when they encounter other agents who do the same
with counter-offers. Agents adjust their offers until an agreement is
reached, after which the transaction is executed by the environment.
When agents were given the ability to drop and pick up items as an
alternative to hand-engineered offers, agents learned to avoid freely
giving away resources, thus necessitating the implementation of a trading
mechanism for exchange to occur. Further experimentation which varied the relative supply and demand of resources showed corresponding changes in price akin to what one might expect from real-world supply/demand curves. It was also shown that under certain maps with apples and bananas on opposite sides, a ``merchant''-like behavior can emerge, where agents trade for apples on the apple-saturated side and then sell them at a higher price on the banana-saturated side.

\subsection{AI Economist}
Aside from Fruit Market, there are other environments that directly implement different mechanisms for exchange. Of note is the AI economist detailed in \citet{zheng_ai_2020}, which tasks agents with gathering wood and stone to construct houses. Agents are set with different skill multipliers such that some agents are able to gather more resources while others make more coins building houses. Additionally, the environment provides a global market to which agents can submit buy and sell orders from anywhere on the map, which are automatically executed once a valid transaction exists. This environment adds an additional 44 actions for trading alone, representing the combination of 11 different price levels, whether the order is a buy or sell, and whether the resource is wood or stone. While adding substantial environmental complexity, the market enables agents to specialize in gathering or building houses and trade for the materials or coins they need. 

\subsection{NeuralMMO}
The exchange system of NeuralMMO described in \citet{suarez_specialization_2022} also introduces a global market with which agents can buy and sell items using gold. Alongside the exchange system, a profession system is introduced which allows agents to produce items needed by other professions. As a result, each profession must purchase items from other professions in order to progress and raise their level. Agents can sell items by posting them to the market along with a price, and agents can simply select an existing item on the marketplace to purchase it at the specified cost. The NeuralMMO exchange system introduces 161 unique item types, making it one of the most complex exchange systems in a multi-agent research environment.

\hypertarget{method}{%
\section{Method}\label{method}}

\hypertarget{multi-agent-reinforcement-learning}{%
\subsection{Multi-Agent Reinforcement
Learning}\label{multi-agent-reinforcement-learning}}

Our environment is represented as a partially observable Markov decision
process described by the tuple \(< S, O, A, P, R, \gamma, N>\). The
observation function \(O\) maps each state \(s \in S\) to the local
observation \(o_t^i\) of the environment at time step $t$ for agent \(i\). The shared
action space between \(N\) agents is denoted by \(A\).  Each agent is
controlled by a policy \(\pi(\theta)\) that is parameterized by a weight
vector \(\theta\). In this setting, agents act one at a time rather than
simultaneously. The probability transition function \(P(s'|s_t^i, a_t^i)\) is
represented by \(P\), where \(a_t^i\) is the action taken by agent \(i\)
at time step $t$ and
and \(s'\) is the new environment state after the action has been taken. Notably, $s'=s_t^{i+1}$ if other agents still need to take their turn for the current time step, otherwise $s'=s_{t+1}^{1}$  The reward function is denoted
by \(R(s_t, a_t)\), and the discount parameter is represented by \(\gamma\).
The objective of each agent $i$ is to maximize its discounted accumulated
reward over an episode of \(T\) time steps, which is represented by
\(\mathbb{E}_{a_t^i,s_t^i}[\Sigma_{t=0}^T \gamma^t R(s_t^i, a_t^i)]\).

\hypertarget{environment}{%
\subsection{Environment}\label{environment}}

\begin{figure}
\hypertarget{fig:env}{%
\centering
\includegraphics[width=0.8\textwidth]{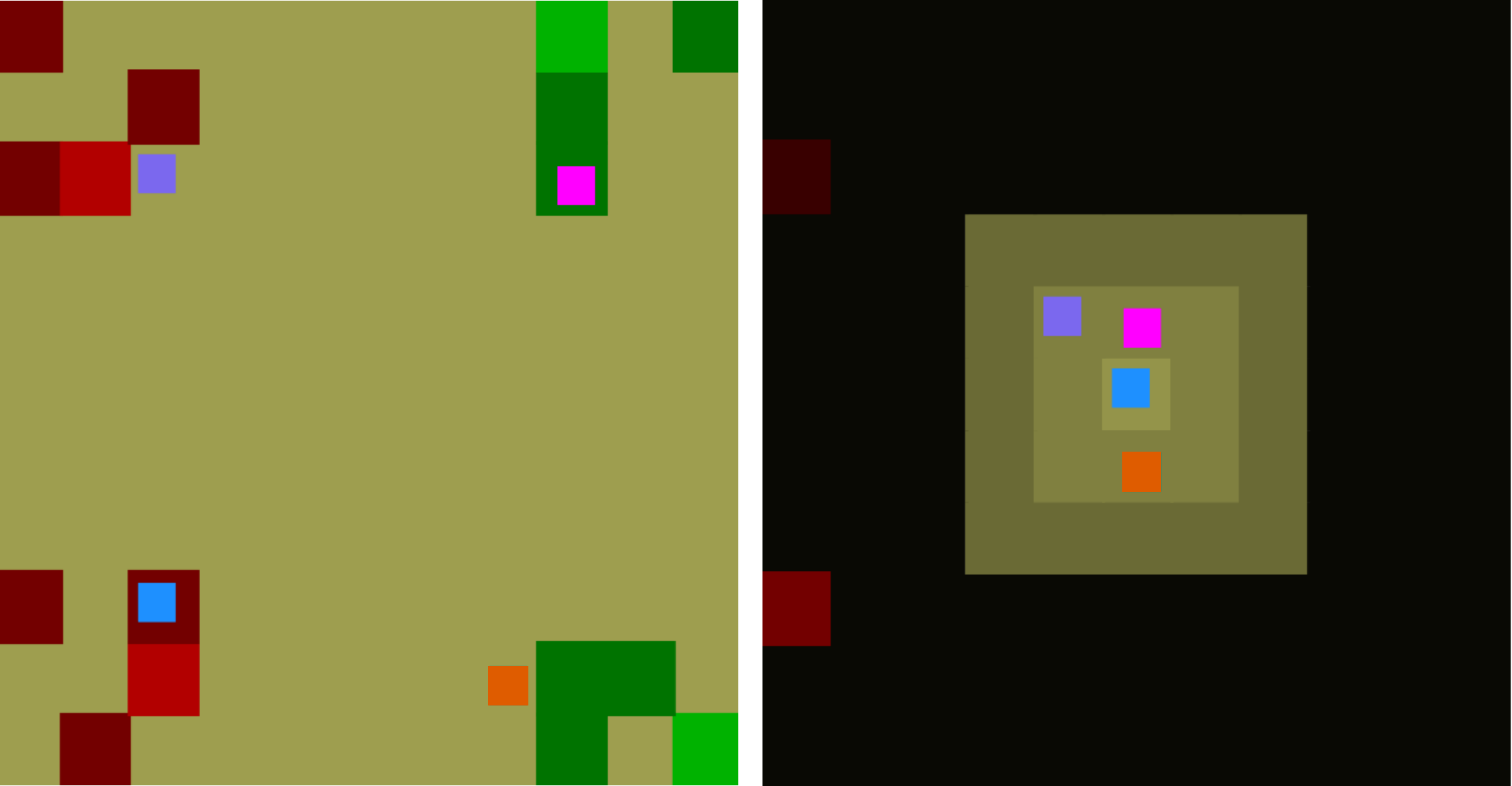}
\caption{\textbf{Left}: The Trading Game environment during the daytime. There are four agents denoted in purple, pink, blue and orange, collecting fruits and greens from patches located in the corners of the grid. Brighter red and green cells indicate more resources are located in those cells. \textbf{Right}: The Trading Game environment during the night time. All four agents are located around a campfire, which provides enough light to avoid the night time penalty in a 3x3 area. The outer 5x5 ring of the fire provides enough light to reduce, but not negate, the light penalty.\label{fig:env}}}
\end{figure}

We formulate our
environment as a two-dimensional grid world with two types of resources, fruits and greens,
denoted by red and green squares respectively. Five fruits spawn randomly in each of the two patches in
 the left corners of the grid and five greens spawn randomly in each of the two patches
in the right corners. Agents are able to move up, down, left, right, or perform no action.
Additionally, agents can pick all fruit or greens on their cell, as well
as place 0.5 fruits or 0.5 greens on their cell, resulting in nine total
actions.

Agents automatically consume 0.1 units of whatever resources they possess
on every time step. If an agent can only consume one type of food, they do not fulfill all their nutritional needs and receive
0.1 reward. If an agent consumes both a fruit and a green in a single
step, then their needs are fulfilled and the agent receives a reward of
1. Thus, in order to maximize reward, agents should consume fruits and
greens together. Unlike \citep{johanson_emergent_2022}, all agents share the
same reward function and are equally proficient at resource collection.
Agents also receive an additional ``collection'' reward equal to the number of newly spawned resources they collected
that time step. Resources placed by other agents do not contribute to the collection reward, as it would be possible generate large amounts of reward by repeatedly placing and picking up resources.

Unlike many environments used to study the emergence of cooperation in
multi-agent systems, our environment has a day/night cycle. The light
level $l$ for each cell on the grid starts at 0, which is the start of a new
day, and then increases to 1 before oscillating between -1 and 1 for the
rest of the episode in small steps. To incentivize agents to avoid dark regions of the map, we introduce a light penalty $p$, which is set to 10 by default. Agents lose $lp$ reward when on cells with a negative light level, which scales the punishment by the darkness of the cell. With the default value of $p=10$, agents can lose up to 10 units of reward in a single step during the middle of the night.  

In order for agents to survive the darkness without receiving
continual punishment, there is a small ``campfire'' region in the center of the
grid which produces a faint light in a 5x5 area. The internal 3x3 area
around it holds a light level greater than zero throughout the entire
episode, and the outer ring holds a light level just under zero. The
addition of the day/night cycle adds an element of periodicity to our
setting; instead of wandering around the entire episode collecting food, we can
expect agent behavior to alternate between foraging during the daytime
and joining up around the campfire at night, treating the campfire like
a ``home base'' described in \citep{isaac_food-sharing_1978}. Agents begin each episode in one of the four corners of the campfire's 3x3 area, spawning in the same corner each episode. At the start of each day, all remaining resources on the grid are removed, and two patches of new fruits and greens spawn randomly around the four corners of
the map.

Days and nights each last 24 time steps, which is enough time for agents to acquire all the resources in a single patch during the day. For an agent to acquire both types of resources on their own, they must stay out during a portion of the night. All experiments shown last 180 time steps which simulates four days of foraging, terminating at midnight on the fourth day. For the purposes of our analysis, we only report exchange statistics from the first three nights, as agents do not always finish trading when the episode terminates in the middle of the fourth night.

\hypertarget{observations}{%
\subsection{Observations}\label{observations}}

Agents observe a local 7x7 area of the grid centered around themselves.
For our setting with 4 agents, each cell contains 18 channels of data,
yielding observation tensors of shape \((7, 7, 18)\). A description of
each channel can be found in Table \ref{tbl:obschannels}.
As agents act sequentially, each observation contains the state of the environment after the previous agent has acted. 

\begin{table}[hbt!]
\centering
\caption{Observation channels. Values for each channel are zero where the description does not specify otherwise. Fruits, greens, and position channels have values only on cells where the is an agent controlled by the respective policy. Notably, since the environment allows agents to share the same policy, we provide a set of ``self'' channels to differentiate an agent between others sharing the same policy. For the experiments shown, this is not needed, as our agents all use a different policy.}
\label{tbl:obschannels}
\begin{tabular}{|l|l|}
\hline
Channel Name & Description \\ \hline
Fruits & \# of fruit on a grid cell \\ \hline
Greens & \# of greens on a grid cell \\ \hline
Light level & Light level for each cell \\ \hline
Self position & 1 where the agent is located \\ \hline
Self Fruits & \# fruit carried by the agent \\ \hline
Self Greens & \# greens carried by the agent on cell \\ \hline
Policy 1 position & \# agents controlled by policy 1 on cell \\ \hline
Policy 1 Fruits & \# fruit carried by agents controlled by policy 1 \\ \hline
Policy 1 Greens & \# green carried by agents controlled by policy 1 \\ \hline
... & ... \\ \hline
Policy 4 position & 1 for cells containing an agent controlled by policy 4 \\ \hline
Policy 4 Fruits & \# fruit carried by agents controlled by policy 4 \\ \hline
Policy 4 Greens & \# green carried by agents controlled by policy 4 \\ \hline
\end{tabular}
\end{table}

\subsection{Algorithm}

We train a deep neural network as our policy using the Proximal Policy
Optimization algorithm (PPO) \citep{schulman_proximal_2017}, leveraging the
implementation provided in the Ray Python library \citep{noauthor_ray_nodate}. While there exist many algorithms
tailored for multi-agent settings
\citep{rashid_qmix_2018, lowe_multi-agent_2020}, vanilla PPO has been
shown to be effective on many multi-agent problems
\citep{yu_surprising_2021,de_witt_is_2020}. All of our agents utilize separate policies which share no parameters. 

Each policy contains a vision network, memory layer, and a controller
network. The output of the vision network is fed into the memory layer,
and the output of the LSTM is sent to the controller, which produces
action probabilities. The architecture details can be found in Table \ref{tbl:arch}, and the full list of
hyperparameters is available in the appendix.

\begin{table}[h]
    \centering
    \caption{Policy Architecture}
    \label{tbl:arch}
    
    \begin{tabular}{|c|c|c|}
        \hline
        \textbf{Sub-module} & \textbf{Layer} & \textbf{Description} \\ \hline
         & 2D Convolution & 128 3x3 filters, padding=1, stride=1\\ \cline{2-3}
        Vision Network & 2D Convolution & 128 3x3 filters, padding=1, stride=1\\ \cline{2-3}
        & 2D Convolution & 128 3x3 filters, padding=1, stride=1\\ \cline{2-3}
        & Flatten & \\  \hline
        Memory Network & LSTM & 512 hidden units \\ \hline
        Controller Network & Dense & 256 hidden units \\ \cline{2-3}
        & Dense & 256 hidden units \\ \hline
    \end{tabular}
\end{table}

\hypertarget{results}{%
\section{Results}\label{results}}

\begin{figure}
\hypertarget{fig:sanity}{%
\centering
\includegraphics[width=1.0\textwidth]{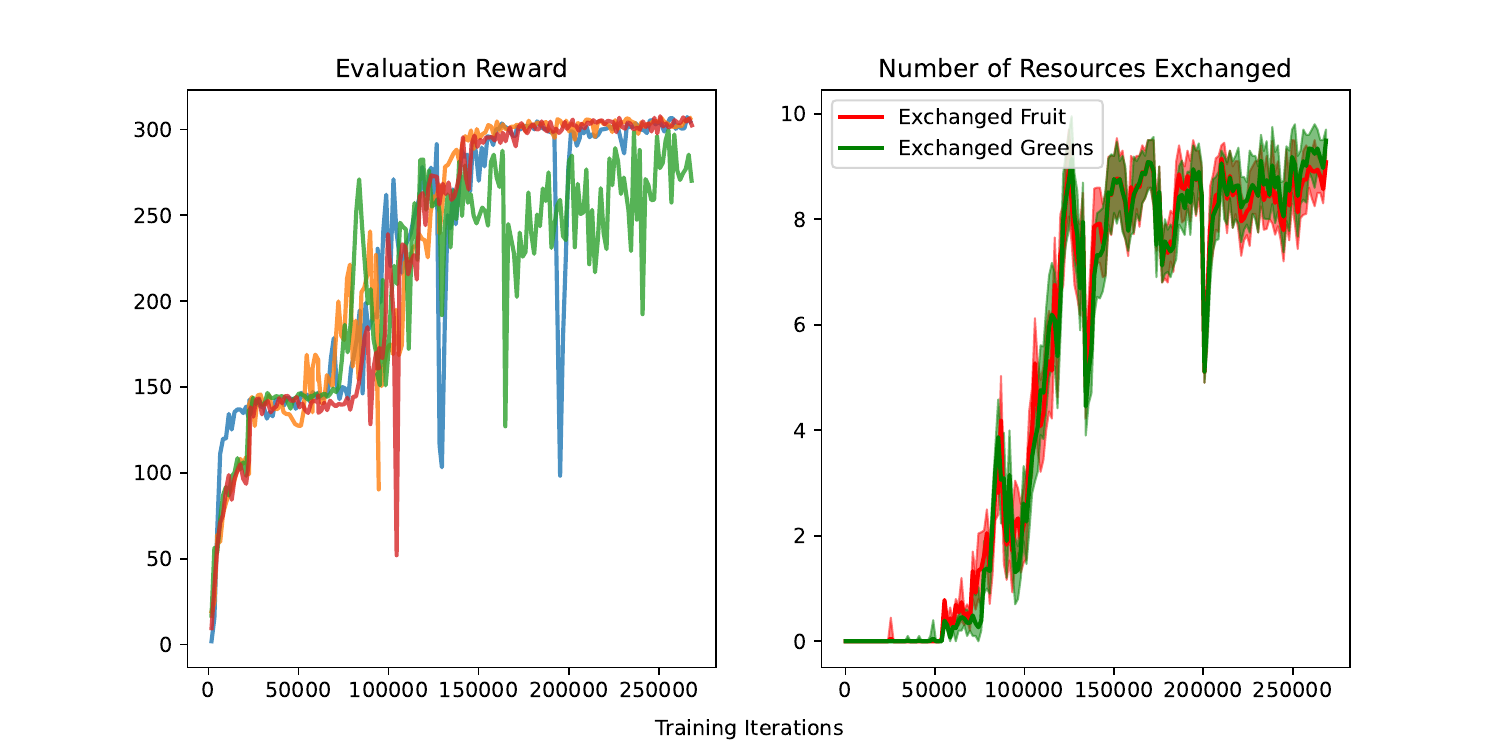}
\caption{Cumulative reward and exchange counts over four trials. \textbf{Left}: Cumulative reward over the duration of a trial, each line represents a one trial. \textbf{Right}: Exchange counts averaged over four trials, standard deviations are shown. The first steps before agents learn to avoid a large night time penalty is omitted for readability in all reward plots. We ran five trials total, but focus on four here as we elect to analyze the fifth trial separately in Section \ref{emergence-of-tolerated-theft}}\label{fig:sanity}
}
\end{figure}

As seen in Figure \ref{fig:sanity}, the behavior throughout training can be
viewed as two periods of equilibrium with a period of fluctuation in
between, reminiscent of the ``punctuated equilibria'' model of
evolution \citep{gould_punctuated_1977}. The first equilibrium is
reached when agents learn to forage resources during the day and return
to the campfire at night, reached at around 10,000 iterations of
training and persisting for around 45,000 more iterations. During this
period, agents do not exchange resources and instead wait out the night
only consuming just the resources they foraged, occasionally dropping a
resource or moving around due to the stochastic sampling of actions
during training.

AI algorithms tend to be employed on games that typically do not contain
long periods of time devoted to doing nothing. Indeed, a game where half
the time is spent doing nothing would likely not be very interesting for many; however, when agents are not given an easy way to further optimize an objective they can find novel, sometimes unexpected methods to do so given
time \citep{baker_emergent_2020}. 

\begin{figure}
\hypertarget{fig:trade}{%
\centering
\includegraphics[width=1.0\textwidth]{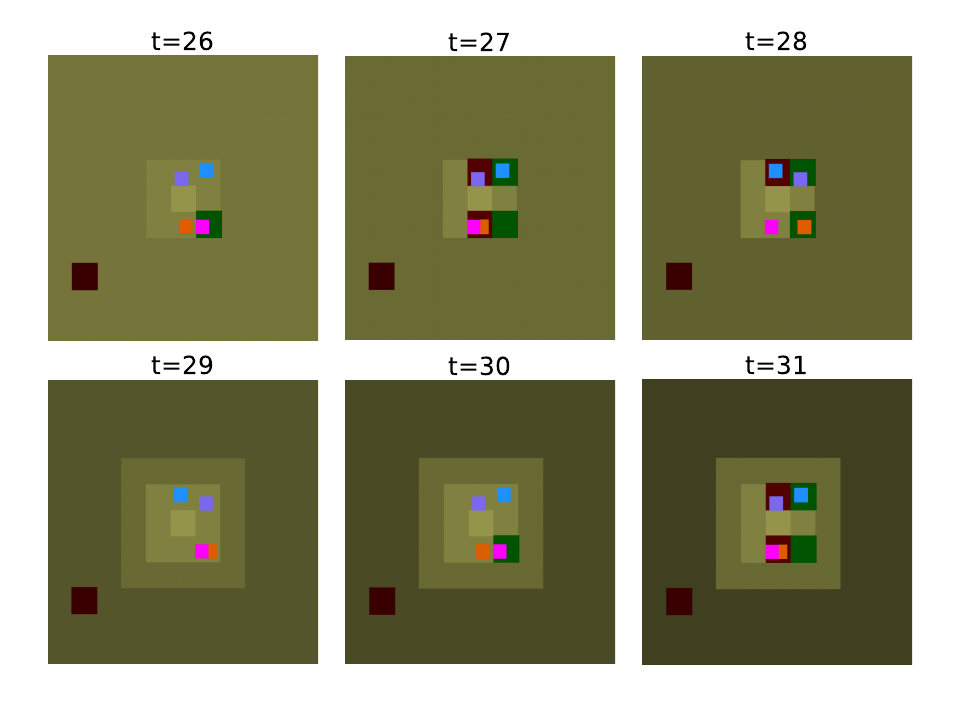}
\caption{Five steps making up an example exchange from top left to
bottom right: Four agents form into two pairs, each agent on an cell
adjacent to their partner. Agents drop half a resource before moving
over to the adjacent cell where they collect the resource dropped by
their partner. After collecting their partner's resource, agents move
back to their original places and begin to perform another
exchange.}\label{fig:trade}
}
\end{figure}

We observe the rise of such novel methods during the
first transition around 55,000 training iterations in, when agents discover the ability to exchange resources around the fire. Exchange starts off in very small quantities, with agents
dropping just half a unit of food in total over three nights, despite
possessing an abundance of their respective resources each night. It
takes thousands more iterations before the number of exchanges over three nights stabilizes around nine fruit and nine greens per episode between
four agents. By this time, agents split up into two pairs who trade with
each other, as seen in Figure \ref{fig:trade}. This averages out to 3 resources
exchanged per night, 1.5 resources per pair of agents. We denote this
trading configuration as 2-PAIR. Each agent goes to one resource patch
per day, and each patch contains 5 units which all get picked. In
the ideal case, each pair of agents would trade 2.5 units; however,
agents need to walk from the patch back to the campfire, consuming
anywhere between 0.5-1.0 units of food in the process. This implies that
that 1.5 resources per night pair of agents is fairly close to the
practical optimal quantity.

An example exchange can be found in Figure \ref{fig:trade}. Agents form into
pairs and stand a cell away from their partner. Each agent drops a
resource before moving over to their partner's cell to collect the
resource dropped by their partner. Notably, the pairs are not adjacent
to each other, which may limit the degree to which different pairs can
interfere with each other.

\hypertarget{emergence-of-tolerated-theft}{%
\subsection{Emergence of Tolerated
Theft}\label{emergence-of-tolerated-theft}}

\begin{figure}
\hypertarget{fig:decoy}{%
\centering
\includegraphics[width=1.0\textwidth]{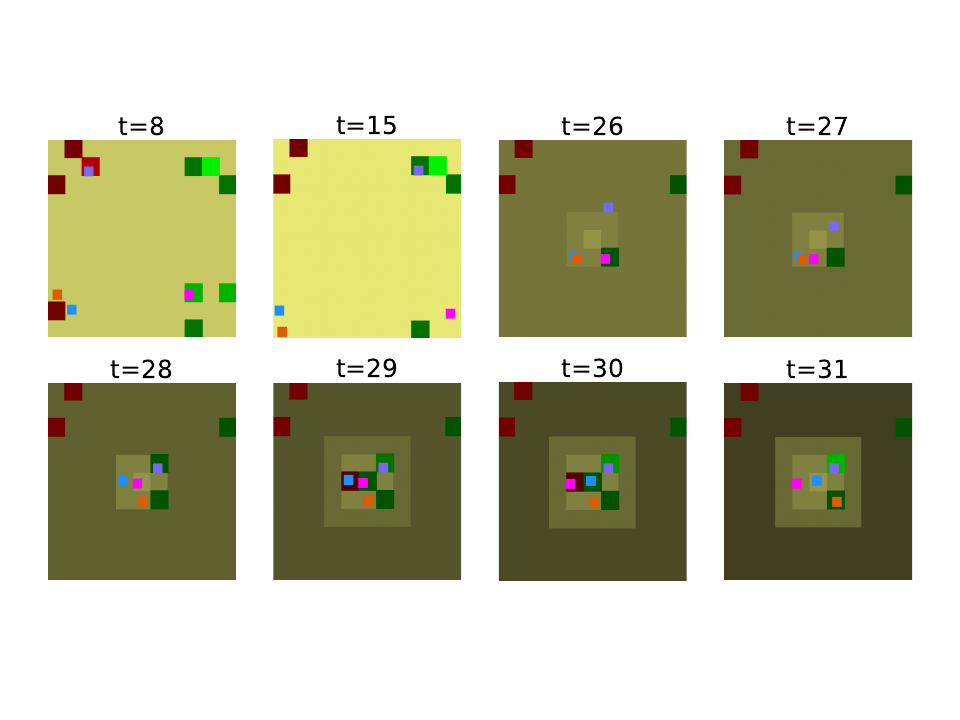}
\caption{Key points during a concession, from top left to bottom right.
\textbf{t=8-15}: Blue and Orange forage the fruits
in the bottom left, Purple goes after the resources at the top
of the map, and the Pink agent gets the entire patch of greens at the
bottom right. \textbf{t=26-27}: Pink drops an offering and
moves towards Blue. \textbf{t=28-29}: Orange moves toward
the offering, allowing Blue and Pink to begin an exchange. Purple
places down greens for no apparent reason, but the greens are not at
risk of immediately getting taken, so this does not negatively affect reward. \textbf{t=30-31}: Orange moves to
collect offering. Blue and Pink finish their
exchange.}\label{fig:decoy}
}
\end{figure}

Across four trials, agents will sort themselves up into pairs to
exchange resources, as reported above. In a fifth trial
however, a different form of behavior emerges which is so interesting that it
deserves its own analysis.

The 2-PAIR trading configuration emerges when each agent finds its own
food patch. It is possible, however, that agents do not divide themselves
evenly across all patches and instead get caught in a local minima where
two agents visit the same patch and split the resources. In this
particular trial, the purple agent collects resources from both a fruit
and a green patch (accepting a minor light penalty in the evening in
order to do so), the blue and orange agent share the other fruit patch,
and the pink agent forages the other greens patch alone. As a result,
the purple has significantly less incentive to trade, leaving the other
three agents to divide foraged resources among each other.

The resulting behavior is fascinating; as seen in Figure \ref{fig:decoy}, the
pink agent will drop some of their excess greens, which lures the orange
agent away from the blue agent. The orange agent then leaves the
blue and pink agents alone to trade and collects their offering
of greens. This behavior is present on every single evaluation run on the final checkpoint
of this trial.

\begin{figure}
\hypertarget{fig:nodecoy}{%
\centering
\includegraphics[width=1.0\textwidth]{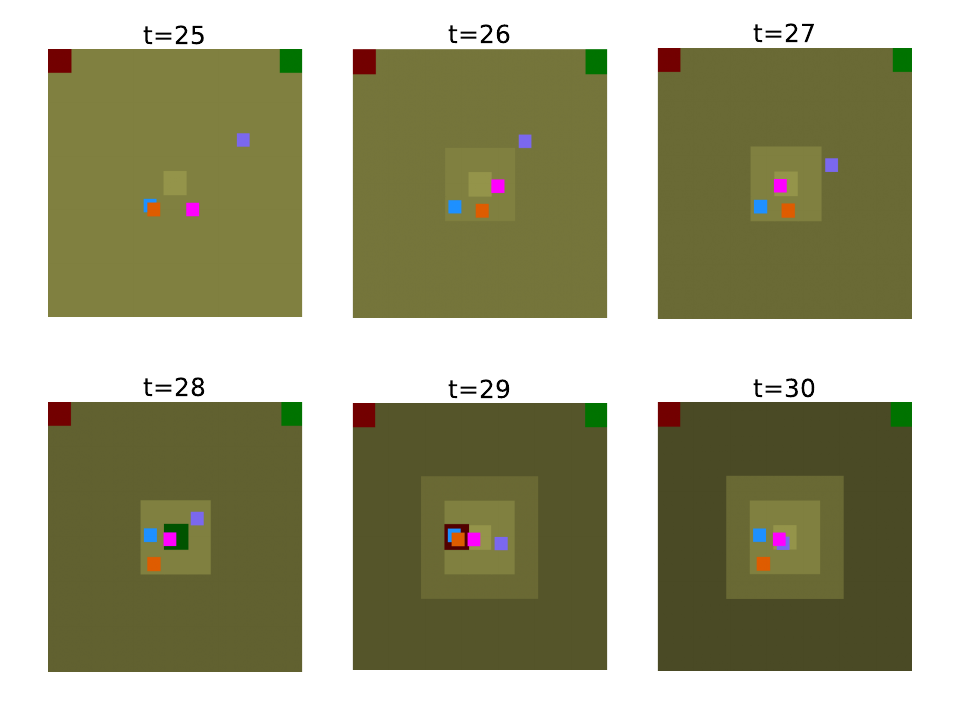}
\caption{Orange interferes with exchange between Pink and Blue
when a concession is not made, from top left to bottom right.
\textbf{t=25-26}: Orange occupies a potential exchange space.
\textbf{t=27-28}: Pink (controlled by a human) goes around Orange
and begins to initiate an exchange with Blue. \textbf{t=29}: Blue offers a fruit for Pink, but Orange moves to
occupy the cell, which allows it to collect the fruit on the next turn
    before Pink can reach it. In response, we (Pink) rescind the offer of greens. \textbf{t=30}: 
 Orange moves away from Blue and Pink, potentially to tempt Blue and Pink into another exchange. }\label{fig:nodecoy}
}
\end{figure}

In order to test whether this behavior is a coincidence or intentional,
we take control of the pink agent to prevent it from dropping the bait,
and observe the change in behavior of the orange agent which can be seen
in Figure \ref{fig:nodecoy}. The orange agent responds by interfering with the
blue and pink agents when they attempt to trade, akin to a
defender in basketball. We control the pink agent during the attempted
exchanges as well, since the pink agent will only attempt to trade with
blue after it drops an offering for orange.
Interestingly, there is no need for the pink agent to wait for a return offer after it has left a resource to orange. This allows the orange agent to collect its resource after the pink agent has moved three cells away to trade with blue, enabling food sharing over a distance of a few cells.

Out of all the theories for the emergence of food sharing, this behavior
is the closest to tolerated theft, where agents freely give resources
because the cost of defending those resources is greater than the cost
of simply giving them away \citep{isaac_food-sharing_1978}. In this case, the cost of defending a resource is replaced with the missed opportunity to exchange. The Trading Game supports no pre-programmed method of punishment like the
punish beam described in \citep{vinitsky_learning_2022}, yet it appears
that the emergence of exchange brings forth both new forms of conflict
(interfering with exchanges) and new ways to deal with troublemakers
(conceding resources).

\hypertarget{resilience-to-defection-and-exploitation-of-suckers}{%
\subsection{Resilience to Defection}\label{resilience-to-defection-and-exploitation-of-suckers}}

In all of the experiments run, agents never exchange resources on the
same grid cell; rather, they consistently stand at least one cell away
from a partner, drop a resource, and if the partner reciprocates, both
agents exchange spots. We call this behavior DROP-SWAP. Notably,
DROP-SWAP is fairly complicated behavior, and we would not expect it to
arise if the only goal between agents was exchange. A much simpler
exchange strategy would involve agents meeting on the same grid cell,
then dropping and picking without moving during the exchange. So why do
partners always keep their distance for each exchange?

\subsubsection{Intra-Pair Defection}

We hypothesized that DROP-SWAP emerges as a mechanism to defend against
defection, since a cooperative agent will have enough time to reclaim
their offer before a defecting partner can grab it. We denote this kind
of defection \emph{intra-pair defection}, since the cheating behavior
takes place within the pair. To test this, we overwrote agent actions
during evaluation to observe how agents respond when a partner reclaims
their offer or attempts to grab an offer without reciprocating.

There are two broad ways an agent can defect, since agents do not perform
actions at the same time. The first defection occurs when an agent
places down a resource, but then picks it back up once their partner
drops before moving to collect the partner's resource. The second
defecting strategy occurs when the partner drops a resource and the
agent attempts to grab it without dropping anything in return.

Across each of the four trials, we overwrote each agent to perform each type of defection during an exchange with their partner. For every single pair
across all trials, the partner defended their resource and rescinded
their offer, moving back to their original cell if needed to rescind. Despite our best efforts, we were unable to trick any partner into giving up their food for free, supporting the hypothesis that the DROP-SWAP form of exchange arises to prevent agents from getting cheated out of their offered resource.

The ubiquity of defense against defection is rather surprising, considering the relative payoffs of getting cheated versus performing an exchange. If an agent is cheated out of half a resource without receiving anything in return, they only lose 0.5 units of total reward. Furthermore, at the time step of the defection, the loss is heavily discounted as it only impacts future reward when agents run out of food five steps sooner. In contrast, performing an exchange yields around an additional 4.5 units of total reward (5 from the exchange - 0.5 from running out of a resource sooner). Reward from exchange is also discounted far less, as agents immediately start receiving the reward for consuming two resources at once. Given these relative payoffs, we might not expect agents to learn to so strongly defend their resource unless it heavily motivated a partner to provide an offer in response.

\subsubsection{Inter-Pair Defection}

Additionally, we noticed that across all trials, pairs never exchange
adjacent to other pairs. Instead, pairs consistently trade with at least
one empty cell in between them. We hypothesized that this might have
emerged as a method to defend against \emph{Inter-pair defection}, where a
pair will refuse to exchange if an outside agent is able to grab the
dropped resources. To test this, we made each agent interfere with the
opposite pair and measured whether it was possible to intercept
resources during the exchange.

We managed to intercept at least
one resource between seven pairs out of eight across four trials. This does not imply, however, that stealing a resource was always easy. we observed various levels of defense: Some
pairs would completely ignore outsiders, while others would refuse to
initiate an exchange if there was an agent adjacent to the pair. In some cases, we could steal a resource from a pair right when they began their first exchange, but after getting cheated, they would refuse to exchange any further.  Notably, a
pair might display more defense against one outsider, but completely
ignore a different outsider and allow them to steal during the exchange.

We speculate that this variability in inter-pair defense may be a result
of the difficulty in discovering exchange. If two pairs discover
exchange at roughly the same time, attempting to intercept resources from
another pair will quickly prove less beneficial than exchanging with a
partner. On the other hand, when one pair discovers exchange far before the other, there
is heavy incentive for outsiders to get better at intercepting resources from
the cooperators until the non-cooperative pair can discover exchange themselves. Under this hypothesis, the difference in time between the two pairs individually discovering exchange may be an indicator to the degree of defense the first pair may have. There is no
clear metric to quantify the degree to which agents defend from inter-pair
defection, so validating this theory is not straightforward.

\hypertarget{sec:intercoop}{%
\subsection{Inter-pair Cooperation}\label{sec:intercoop}}

When exchanging resources in trials without tolerated theft, agents form
into pairs and exchange resources with their partner. As each agent is
perceived in a separate observation channel, we sought to understand the
degree to which this exchange behavior is tied to a particular partner.

\begin{figure}
{%
\centering
\includegraphics[width=0.8\textwidth]{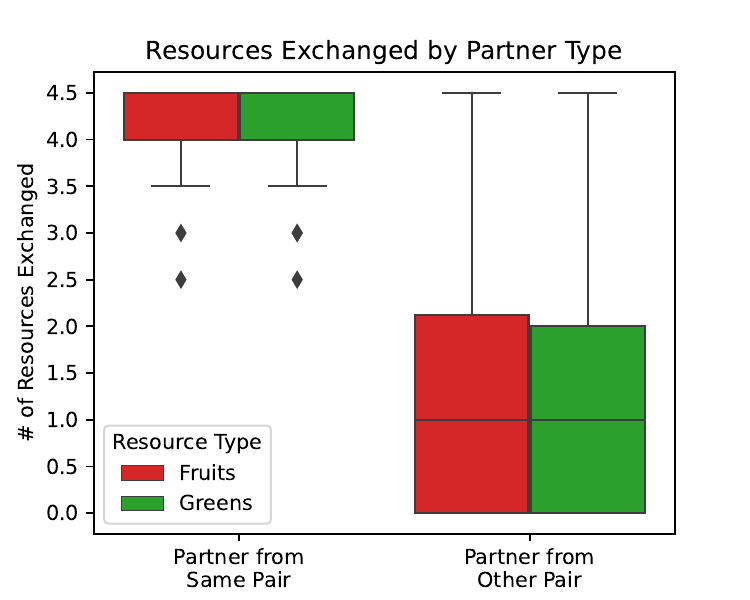}
\caption{Number of resources exchanged over three nights between two agents. Results shown are taken from ten evaluations for two partner pairings across four trials, yielding 80 data points per distribution. We expect optimal agents to exchange 4.5 units
of a given resource over three nights, which is approximately reached
when agents trade with their preferred partner. When trading with
another agent's partner, rates of exchange vary
wildly.}\label{fig:inter}
}
\end{figure}

We measured this by freezing the normal partners from entering the
campfire, and seeing if agents from opposite pairs would exchange
resources around the campfire if their usual partner was not available.
The exchange counts over three nights can be found in Figure \ref{fig:inter}.

As in the defection experiment, the results were varied: many inter-pair
pairings exchanged no resources, some pairings exchanged the near
optimal amount, and other pairings would only exchange half a resource.
These results imply that, despite extensive periods of time to explore
interactions with other agents around the campfire, the degree to which
agents explore the full range of interactions with others varies greatly.

\hypertarget{different-quantities}{%
\subsection{Different quantities}\label{different-quantities}}

In Figure \ref{fig:sanity}, we see that there is approximately a 1:1 exchange
ratio between fruits and greens, which logically follows from the 1:1
distribution of resources. This prompts the following question: How do the
rates of exchange change in settings with asymmetrical distributions of
resources?

We ran two groups of five trials to answer this question. Tolerated theft emerges once in the first group and twice in the second group, so we focus on three
trials from each group that do not demonstrate tolerated theft, as tolerated theft skews base exchange rate away from 1:1. For one
group, we set the fruit and greens patches to contain six fruits and
four greens respectively. The other group had six fruits in their fruit
patches and four greens in their greens patches.

\begin{figure}
\centering
\includegraphics[width=1.0\textwidth]{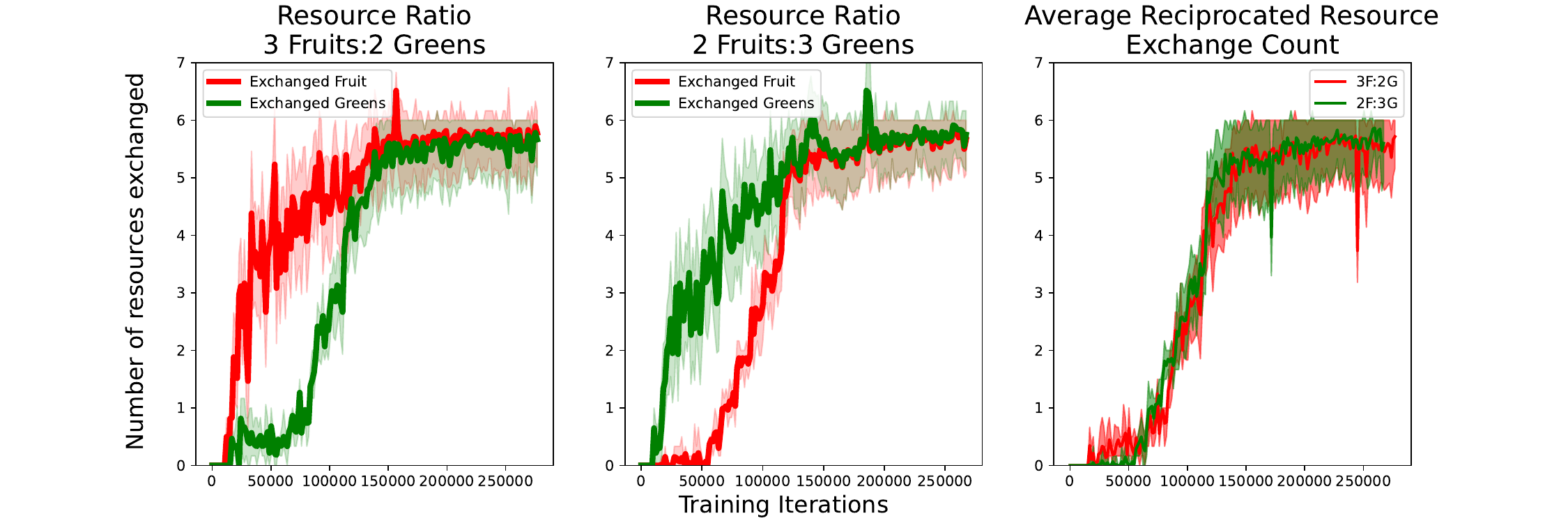}
\caption{Exchange quantities over three nights on maps with different
quantities of resources, over three trials. Standard deviations are
shown. \textbf{Left}: Fruit patches in this map contain six fruit, while
the patches of greens contain four greens. \textbf{Middle}: Patches of
greens contain six greens, while fruit patches contain four fruit. \textbf{Right}: The average number of resources reciprocated, e.g., if agent $a$ gives four fruits to $b$, but $b$ only gives two greens to $a$, then two resources were reciprocated. Where the abundant resource is initially given for nothing in return, the scarcer resource is only exchanged as a form of reciprocation.}
\label{fig:diffquantsfig}
\end{figure}

The number of exchanges for each resource can be found in Figure \ref{fig:diffquantsfig}, where we can observe interesting dynamics play out over over the course of training. Initially, agents are
willing to give great amounts of the abundant resource, often for nothing in return. In this setting, there is enough of the abundant resource for an agent to never run out. When an agent has so much food that it will never go a step without at least one resource to consume, dropping some of the extra resources yields the same reward as hoarding resources that will never be eaten. Thus, agents with excess will occasionally drop their food, since there is no reason not to. To get agents to drop more than a spare resource here and there, some extra reward for doing so is required. Agents with the scarcer resource begin to offer food which provides that reward, and so DROP-SWAP emerges. Eventually, the exchange rate approximates 1:1., with the abundant resource exchanged in slightly
greater quantities than the scarcer resource.

The approach towards a 1:1 ratio is surprising. Given the 3:2
distribution of the resources, we might intuitively expect a 3:2
exchange rate to stabilize, as was the case in
\citep{johanson_emergent_2022}, but agents appear to stabilize on the
DROP-SWAP strategy in a 1:1 ratio, despite initially giving resources
away for free and possessing the ability to perform a 3:2 exchange in a
single transaction. Nevertheless, after all 1:1 exchanges take place for
a night, agents with the abundant resource consume whatever leftovers
they have, and the cycle repeats the next day. For now, we do not
attempt to provide an explanation for this phenomenon, and instead leave
in-depth study of this dynamic to future work.

\hypertarget{ablation-studies}{%
\subsection{Lowered Light Penalty}\label{ablation-studies}}

In the Trading Game, the night penalty $p$ serves multiple
purposes: 1) to pressure agents into congregating for extended periods of
time and 2) to prevent agents from foraging from both sides of the map
in the same day. With the rather high value $p=10$, we've seen that agents will forage a single resource, then exchange for the other food type around the campfire; with a lower $p$, we might expect agents to stay out during parts of the night to forage both resources for themselves, as the weaker night penalty no longer outweighs the extra benefits from staying out to collect the other resource. Given this, we can view $p$ as a parameter which controls the duration and degree to which agents will congregate around the fire.

\begin{figure}
\centering
\includegraphics[width=1.0\textwidth]{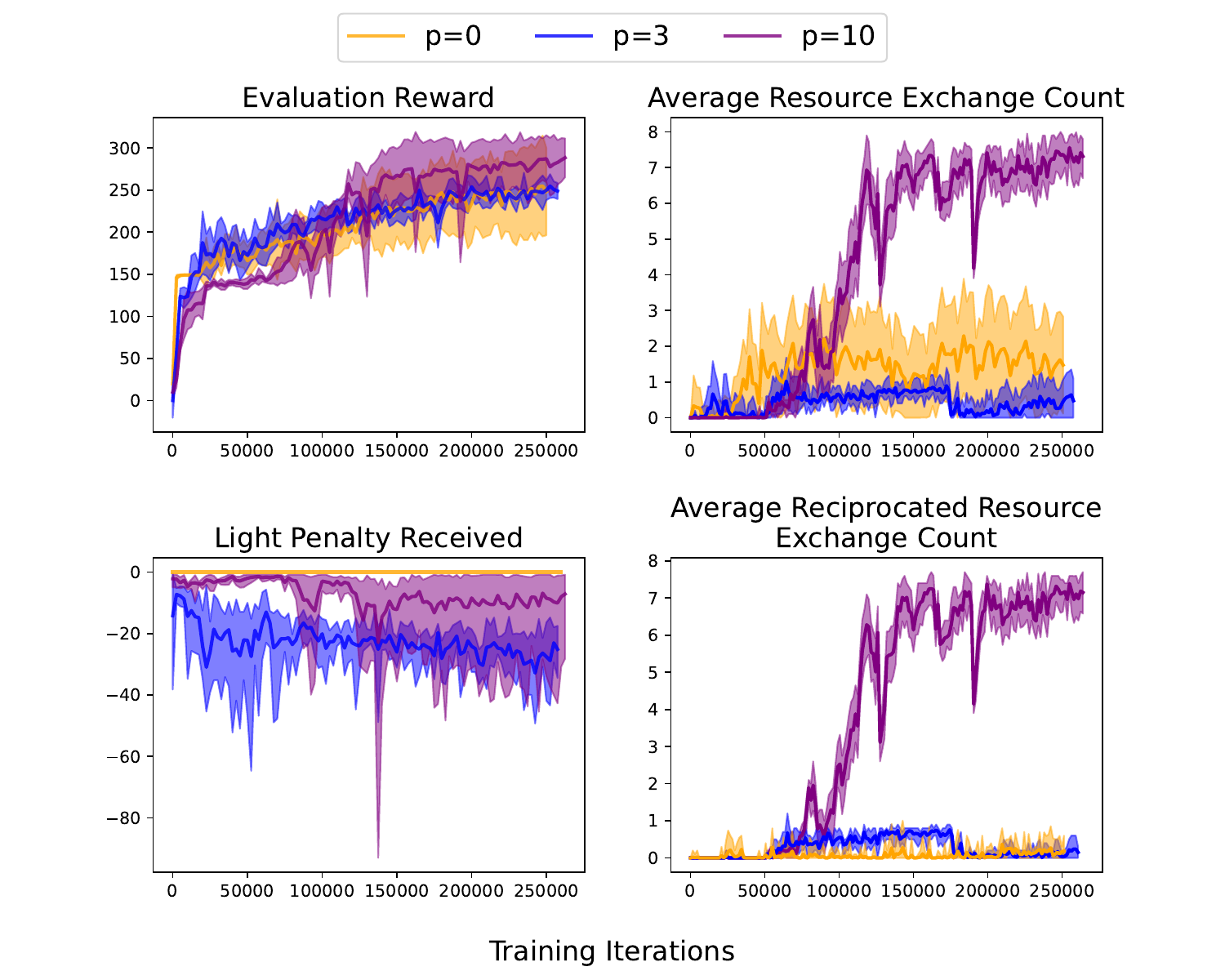}
\caption{Metrics for $p=0, p=3$ and $p=10$ over five trials each, including tolerated thefts. Standard deviations are shown.\textbf{Top Left}: Total evaluation reward. Notably, agents that receive a larger light punishment achieve higher collective reward than groups of agents who receive a weak light penalty and groups of agents who receive no penalty. \textbf{Top Right}: Exchange counts with different light penalties over five trials, standard deviations are shown. We observe that with a lower light penalty, agents exchange less, as there is more incentive for an agent to stay out at night to collect both kinds of resources on their own. Counter-intuitively, more exchanges occur when there is no light penalty than when there is a weak one. \textbf{Bottom Left}: When $p=3$, agents stay out during the night longer than when $p=10$, and thus receive a greater total punishment over time. \textbf{Bottom Right}: The number of reciprocated resources between two agents. Where we see a modest amount of exchange occur when $p=0$, we see here that very little of it is reciprocated. \label{fig:light}} 
\end{figure}

To study the impact of the light penalty, we run two sets of five trials, one where $p=0$ and the other where $p=3$, and compare them to the five $p=10$ trials. The rewards and exchange counts for these trials can be found in Figure \ref{fig:light}. We notice that agents in the lower-penalty setting exchange less resources, receive less total reward, and accept a larger cumulative light penalty than agents from the $p=10$ run. Despite possessing the capability for resource exchange, agents in the lower-penalty setting converge to the aforementioned minima and seem to be unable to escape this minima once it is reached. This emphasizes just how unlikely resource exchange is to dominate as a strategy in the Trading Game: not only must exchange be more rewarding than foraging both resources alone, but foraging both resources needs to be more costly than only foraging one. 

When there is no light penalty and thus no reason to gather around the campfire, we observe slightly more exchange than in the low-penalty case. The exchanges seen here are a result of agents dropping excess resources that they will never consume, as described in Section \ref{different-quantities}. While there are no asymmetries in the quantities of resources, it is possible for agents to forage both patches of a resource and collect more of one food type than they will eat. These agents will occasionally drop their excess as there is no benefit to hoarding, allowing a one-way exchange to occur. This is made clear by the lack of reciprocation occurring during these trials.

In trials with no light penalty, we observed numerous opportunities where two nearby agents holding different resources could perform DROP-SWAP, but opted not to. This cements the importance of the campfire: In Section \ref{different-quantities}, we showed when one agent is willing to give away resources for free, reciprocal exchange can emerge around the campfire to further promote exchange. Without a light penalty (and thus no extended downtime around the campfire) some agents give away resources for free, but nobody reciprocates,  and thus DROP-SWAP fails to emerge. 

\hypertarget{discussion}{%
\section{Discussion}\label{discussion}}

This work takes inspiration from the concepts of autocurricula from reinforcement learning \citep{leibo_autocurricula_2019} and coevolutionary arms races \citep{ficici_challenges_1998}, where agents create problems for others to solve, which then leads to the creation of clever solutions and even harder problems. 
In our domain, these dynamics produce the emergence of exchange, the emergence of
tolerated theft-like behavior, and competitive-cooperative dynamics such as defending exchange offers
from defectors. Notably, unlike all exchange systems from previous work, the complexity of this environment arises not
from agents learning to use complex game mechanics with many
actions, but rather by learning complex ways to use just nine. Given the actions of picking up and placing down resources, agents exhibit complex forms of cooperation and competition in ways we have not intended or expected. If the environment had some human-designed trading system which allowed no room for interference or defection, these dynamics likely would have not arisen.

The campfire does not explicitly facilitate
exchange, but instead acts as a stepping stone \citep{stanley_why_2015}
for its emergence by shaping the conditions under which interactions occur. In the work presented, it takes up to four days of wall clock time for agents to begin to reliably exchange resources on a domain where approximately 50\% of the training steps are spent in a 3x3 area with four other agents. One need not imagine how long it might take for these behaviors to emerge in the $p=0$ setting, where the chances of interacting with another agent are significantly lowered--ten days of real-world training time proved insufficient. Concepts like the ``punish beam'' from \citep{vinitsky_learning_2022} or the broadcast radius of trade offers \citep{johanson_emergent_2022} can also increase the chance of interacting with an other agent, mitigating this effect, but the campfire is unique in that it emphasizes repeated interactions with the same partner during training under similar conditions. Furthermore, the campfire does not require all actions to have a large range or area of effect, allowing exchange to emerge without requiring drop/pickup actions to apply over a distance, as seen with the emergence of tolerated theft.

While the Trading Game leverages a heavy light penalty and distant resource piles to prevent agents from initially foraging both resources, the main purpose of the campfire is to promote periods of extended, idle congregation. There are a plethora of environmental modifications which would disincentivize dual foraging, such as adding difficult terrain between resources for which agents incur a penalty to cross or making agents proficient at foraging different resource types. The goal of this work is not then to present the Trading Game as some benchmark task that should be accepted ``as-is'', but instead as a set of ideas that can be incorporated and modified in other environments to study new forms of cooperation.

\hypertarget{sec:ipd}{%
\subsection{Informal Reduction to Iterated Prisoner's Dilemma}\label{sec:ipd}}

Informally, the emergence of resource exchange during the night can be
reasoned about in a similar fashion to the emergence of cooperation
in the iterated prisoner's dilemma (IPD). The night reduces the practical
dimensionality of the environment, pushing agents to the small 3x3
campfire to perform whatever actions they please for the duration of the
night. This setting mirrors the IPD, where agents play multiple rounds with
the same player; if agents could roam around as they pleased during the
entire episode, random interactions between agents would be rare and
unlikely to be iterated if agents move apart after the interaction. This
may explain why resource exchange did not occur in the Fruit Market
environment: A form of cooperation like modern-day markets may not need
repeated interactions with the same partner, but for a behavior like resource exchange to arise,
repeated interactions with previously seen individuals might be a necessary stepping stone
\citep{henrich_weirdest_2021, stanley_why_2015}.

We expect selfish agents to first learn to avoid dropping resources
around the fire, and pick up any resources that are dropped by others.
This simple short-term method of maximizing reward is analogous to defection in the prisoner's dilemma. This is
indeed the first equilibrium we observe, and it persists for many
thousands of training iterations.

Due to the stochastic sampling of actions from our policies during
training, agents still occasionally drop their resources around the
fire, enabling agents to accidentally gift each other resources for
extra reward. Agents which drop resources without receiving anything in
return will avoid dropping resources in the future, which implies that
if agent \(a\) wants agent \(b\) to drop a resource more often, it needs
to make dropping a resource provide a higher reward for \(b\) than not
dropping a resource, which it can do by offering or not offering a
resource in return. This can be thought of as analogous to Tit-For-Tat,
where defection and cooperation are both reciprocated such that
cooperative strategies receive a higher reward on average than defecting
strategies.

Like the noisy iterated prisoner's dilemma described in
\citep{lindren_evolutionary_1992}, there are no clever agents which
intentionally shape the behavior of other agents. Rather, new forms of cooperation (like exchange) lead to new forms of defection (cheating/interfering), which leads to more complex forms of cooperation (tolerated theft/defending offers). Defection is not a viable long-term strategy since agents alter their policies to respond to defection, so when
this defection-only minima is escaped, it must be from a cooperative strategy which is resistant to exploitation. This is likely why we always observe the rise of DROP-SWAP trading across all trials which converge to 2-PAIR.

\hypertarget{limitations}{%
\subsection{Limitations}\label{limitations}}

As seen in Section \ref{sec:intercoop}, while agents can reliably exchange with
their partner of choice, the degree to which this behavior generalizes
to other agents varies. Furthermore, unlike Fruit Market and the AI Economist, exchange in
the Trading Game emerges when agents are unable to acquire both
resources on their own without getting heavily penalized and are required to work together to maximize reward. This is not uncommon, however, as reference games which require cooperation to solve are used to study the emergence of natural language \citep{lazaridou_multi-agent_2017}.

This domain is also sensitive to the quantity of resources. If there are too few fruits, agents will consume them before they get a chance to trade; too many fruits, and agents can forage fruits one day and greens the following day, as there will be leftovers from the night before, thus reducing the benefit of mutual exchange. Furthermore, resources need to be spaced out: if fruits and greens spawned next to each other, there would be far less reason to specialize and exchange. These pitfalls could be alleviated by adding a limit to the amount of resources an agent can carry at once, or by adding incentive for agents to specialize in a particular resource as done in \citep{johanson_emergent_2022}, but for the purposes of this work we keep the environment as simple as possible.

The compute required to reach exchange emergence is also nontrivial, requiring up to four days of wall-clock time on a Titan X before a pair may begin to reliably trade, and up to ten days in experimental settings with eight agents. This bottle-necked iteration speed and made it difficult to predict when an experimental configuration could yield emergent exchange or not, as performance does not improve significantly during the first equilibrium. Environments supporting greater social complexity with larger numbers of agents would likely take significantly longer.

\hypertarget{futurework}{%
\subsection{Future Work}\label{futurework}}
Given the results and limitations described in the Trading Game, there remains ample room for future work. 
Within the Trading Game, there still exists many interesting environmental properties to study, such as the potential effects of adding a carrying capacity for resources or relative food quantities.
Generalizing this exchange behavior to all seen and even unseen agents is also of great interest. 

The simplicity of the Trading Game also makes it ripe for extension; for example, the addition of a local communication system could allow agents to negotiate around the campfire.
Deep neuroevolutionary approaches have been successfully applied as competitive alternatives to single and multi-agent reinforcement learning problems \citep{such_deep_2018, klijn_coevolutionary_2021} and show potential as another algorithm for the study of emergent cooperation.
Concepts analogous to the campfire could be applied to related domains like Fruit Market and even to entirely different domains such as reference games for research on emergent communication \citep{lazaridou_emergent_2020}.

\hypertarget{conclusion}{%
\section{Conclusion}\label{conclusion}}

In this work, we demonstrated how novel behaviors can arise by reshaping the environmental conditions under which agents interact.
We showed that a simple foraging environment with periodic gathering around a campfire can lead to emergent trading behavior, and discussed how the emergence of trading in our setting is analogous to the evolution of cooperation in the Iterated Prisoner's Dilemma. By directly interacting with the agents, we found that agents could prevent themselves from being cheated by their usual partner, but they exhibited varying levels of defense against being cheated by a third party. 
Additionally, we observed that agents could learn to interfere with exchanges as an indirect form of punishment, allowing an emergent behavior similar to tolerated theft to emerge. As congregation pressure is reduced, these behaviors arise in much weaker forms, if at all, demonstrating the importance of extended congregation on the emergence of embodied cooperation.

\bibliography{NAME}

\hypertarget{appendix}{%
\section{Appendix}\label{appendix}}

We use RLlib provided in Ray 1.13.0 \citep{noauthor_ray_nodate} and
PyTorch 1.11.0 in our experiments. All parameters not mentioned in Table \ref{tab:ppoparam}
can be assumed to be the default for these versions.

\begin{table}[h]
    \centering
    \begin{tabular}{|c|c|}
        \hline
        Hyperparameter Name & Value \\
        \hline
        Learning Rate & 1e-4 \\
        Gamma & 0.99 \\
        GAE $\lambda$ & 0.95 \\
        Clip range & 0.03 \\
        Entropy coefficient & 0.05 \\
        Value Function Coefficient & 0.25 \\
        Num SGD Iterations & 5 \\
        Batch size & 2000 \\
        Minibatch size & 2000 \\
        LSTM Max Sequence Length & 50 \\
        \hline
    \end{tabular}
    \caption{PPO Hyperparameters \label{tab:ppoparam}}
\end{table}

\end{document}